\documentclass[12pt]{article}
\usepackage{epsfig}
\usepackage{amssymb, amsmath}
\usepackage{sint_modi,cite}

\usepackage{graphics}
\usepackage{graphicx}
\usepackage{amssymb}
\usepackage{epsfig}

\def\a{\alpha}
\def\b{\beta}
\def\g{\gamma}

\def\e{\varepsilon}

\def\m{{\tt -}}

\def\r{\rangle}

\def\xprim2bar{\overline{x}^{\prime\prime}}

\newcommand{\beqa}{\begin{eqnarray}}
\newcommand{\eeqa}{\end{eqnarray}}

\let\a=\alpha   \let\b=\beta   \let\g=\gamma   
\let\e=\epsilon         
        \let\m=\mu
                 \let\r=\rho

     \let\L=\Lambda

\let\a=\alpha   \let\b=\beta   \let\g=\gamma   
\let\e=\epsilon         
        \let\m=\mu
                 \let\r=\rho

     \let\L=\Lambda

\newcommand{\be}{\begin{equation}}
\newcommand{\ee}{\end{equation}}
\newcommand{\I}{{\rm I_0^4}}
\newcommand{\dl} {{\rm d}l}
\newcommand{\dt}{{\rm d} t}
\newcommand{\half}{\frac{1}{2}}

\newcommand{\bea}{\begin{eqnarray}}
\newcommand{\eea}{\end{eqnarray}}

\usepackage{graphics}
\def\tr{{\rm tr}}
\newcommand{\eq}[1]{Eq.~(\ref{#1})}
\newcommand{\eqs}[2]{Eqs.~(\ref{#1}) and (\ref{#2})}
\newcommand{\fig}[1]{Fig.~\ref{#1}}
\newcommand{\tab}[1]{Table~\ref{#1}}

\begin{document}
%%%%%%%%%%%%%%%%%%%%%%%%%

\begin{titlepage}
\setcounter{page}{1}
\rightline{WUB/07-06}
\vspace{2.0cm}
\begin{center}{\Large \bf The Higgs mechanism as a cut-off effect}
\vspace{.17in}

\vspace{2cm}

{\large Nikos Irges$^1$, Francesco Knechtli$^2$ and Magdalena Luz$^2$}\\ 
\vspace{.12in}
1. {\it High Energy and Elementary Particle Physics Division \\
Department of Physics, University of Crete, 71003 Heraklion, Greece\\
e-mail: {\tt irges@physics.uoc.gr}\\} 
\vspace{.12in}
2. {\it  Fachbereich C, Bergische Universit{\"a}t Wuppertal\\ 42097 Wuppertal, Germany\\
e-mail:\\ {\tt knechtli@physik.uni-wuppertal.de}, {\tt luz@physik.uni-wuppertal.de}}\\

\end{center}
\vspace{3cm}

\begin{abstract}
We compute the Coleman--Weinberg potential with a finite cut-off for 
pure $SU(2)$ and $SU(3)$ five-dimensional gauge theories compactified on an interval.
We show that besides the expected Coulomb phase located at and in the vicinity
of the free infrared stable or "trivial" fixed point, the theory possesses also a
Higgs phase. We compare the results from the potential computation with lattice data
from simulations.

\end{abstract}
\smallskip
\end{titlepage}

\newpage

%%%%%%%%%%%%%%%%%%%%%%%%%%%%
\section{Introduction}

A five-dimensional pure $SU(N)$ gauge theory is "trivial". 
This remains true when one of the dimensions is compactified on a circle of
radius $R$ or an interval. 
The simplest way to see this is to look at the two dimensionless
quantities parameterizing the theory
\bea
 N_5\;=\;\pi R\Lambda \quad\mbox{and}\quad \beta=2N/(g_5^2\Lambda) \,.
 \label{paramspace}
\eea
$N_5$ is the ratio of the cut-off $\Lambda$ over the
compactification scale $1/(\pi R)$ and
the coupling $\beta$ is derived from the dimensionful five-dimensional
gauge coupling $g_5$ (and is appropriate for a lattice cut-off as we will use
in this work).
The crucial observation is that at a critical value of the coupling
$\beta=\beta_c$, a first order phase transition occurs
\cite{Creutz:1979dw,Beard:1997ic,Ejiri:2000fc,Farakos:2002zb}.
For $\beta<\beta_c$ the system finds itself in a confined phase and for
$\beta>\beta_c$ in a deconfined phase.
The non-perturbative spectrum of a pure $SU(2)$ five-dimensional
gauge theory compactified on an interval \cite{Irges:2004gy,Knechtli:2005dw}
was determined by lattice simulations
\cite{Irges:2006zf,Irges:2006hg}. The outcome is that the particle masses
in lattice units are presumably too large in the confined phase in order
to be ``measured'' in simulations. The situation is different in the
deconfined phase, where we therefore expect to possibly reproduce the
Higgs sector of the Standard Model.
Because of these facts, removing the cut-off
(i.e. taking the limit $\Lambda\rightarrow\infty$) 
in the deconfined phase, while keeping the dimensionless coupling 
$g_5^2\Lambda$ perturbative in the limit, is possible only 
at the trivial point --- $g_5\rightarrow 0$ and $\b\rightarrow \infty$ ---
of the phase diagram (the $N_5$--$\beta$ plane).
For an extra dimension of non-zero size\footnote{
We will avoid at all times the singular limit $R=0$.}, in this limit, also 
$N_5\rightarrow \infty$. 
Perturbative computations performed with an infinite cut-off
amount therefore to being located at the trivial point of the 
phase diagram, where interactions vanish.
Any higher dimensional gauge theory is expected to have these generic
properties \cite{ZinnJustin-Book}. 

More precisely, triviality can be understood by looking at the effective
four-dimensional coupling
\bea
 g_4^2 & = & \frac{g_5^2}{2\pi R}=\frac{N}{N_5\b} \,. \label{4dgcoupling}
\eea
The evolution of this coupling with the scale is a calculation that requires a
cut-off. This was done in \cite{Dienes:1998vg}, where the following 1-loop
formula was found for one extra dimension
\bea
(\beta N_5) (\Lambda) & = & (\beta N_5) (\mu) 
- b_4\ln{(\Lambda/\mu)} +b_5 [ (\Lambda/\mu) - 1 ] \,, \label{1looprunning}
\eea
expressed here in the dimensionless parameters using \eq{4dgcoupling}.
$\mu$ is a reference scale such that $1/R < \mu < \Lambda$ and
$R$ is kept fixed.
$b_4$ is a model dependent number and $b_5$ is also cut-off independent 
in the limit of large $\Lambda/\mu$.
One immediately sees from \eq{1looprunning} that as
$\Lambda\rightarrow\infty$, the 1-loop corrected effective coupling $g_4$
goes to zero.
For large $N_5$ and $\beta$ the formula
describes physics in the vicinity of the trivial point,
while its applicability becomes 
questionable as $N_5 \beta\rightarrow 0$ where cut-off effects run
out of control.
Nevertheless it is legitimate to ask how much one can really lower the
product $N_5 \beta$ so that the theory can be safely described analytically
while keeping $g_4^2$ perturbative. 

Our goal is to study spontaneous symmetry breaking (SSB).
Gauge coupling evolution is not the most appropriate tool, since
SSB has to be put by hand in the $\beta$-function computation. 
Instead, a scalar potential is needed.
In particular, one-loop computations of the Coleman--Weinberg potential
\cite{Coleman:1973jx} for compactified extra-dimensional gauge theories
have been carried out at infinite cut-off aiming to explain the Higgs mechanism. 
According to this scenario, (some of) the fifth dimensional components 
of the gauge fields play the role of the Higgs field \cite{Forgacs:1979zs,Hosotani:1983xw,Hosotani:1989bm,Sakamura:2006rf,Hosotani:2007mn,Hosotani:2007qw,Antoniadis:2001cv,Kubo:2001zc}.
Several interesting properties have emerged, some of them encouraging some
of them not from a phenomenological point of view.
The most impressive virtue is the finiteness of the Higgs mass, which is
believed to hold non-perturbatively due to the non-local origin of the
operator whose fluctuations are responsible for generating this mass.
For the same reason it seems though impossible for the rank of the bulk
gauge symmetry to be broken.
To break the rank and/or to obtain reasonable phenomenology,  
additional assumptions have to be employed.
One common feature of these models is the
introduction of extra matter fields, as SSB seems not to be possible
in the pure gauge theory.
Recent works related to such issues 
include \cite{Csaki:2002ur,Haba:2002py,delAguila:2003bh,Scrucca:2003ra,
Csaki:2003dt,Biggio:2004kr,Panico:2005ft,Cacciapaglia:2005da,Panico:2005dh,
Maru:2006wx,Grojean:2006bp,Lim:2007ea,Giudice:2007fh,Agashe:2007mc,
Sakamura:2007qz,vonGersdorff:2007uz,Barbieri:2007bh,Medina:2007hz,Lim:2007jv}.
In any case it seems rather hard to obtain naturally a reasonable hierarchy
of masses in the sector which is supposed to reproduce physics in the bosonic sector of
the Standard Model.

In this work we intend to show that in the interior of the $N_5$--$\beta$ phase
diagram where the cut-off is finite, there is a transition into a broken rank phase of the five-dimensional pure $SU(N)$ gauge theory.
The tool that allows to see this is the
(not necessarily perturbative in $\beta$) expansion in the cut-off 
\bea
-{\cal L} & = & \frac{1}{2g_5^2}{\rm tr}\{F\cdot F\}+
\sum_{p_i}{c^{(p_i)}(N_5,\beta)}\;{a^{p_i-4}}\;{\cal O}^{(p_i)}+\ldots
\label{efflagran}
\eea
of the effective lagrangean
\cite{Symanzik:1981hc,Symanzik:1983dc,Symanzik:1983gh,Luscher:1998pe}, 
where $F$ is the field strength, ${\cal O}^{(p_i)}$ is an operator\footnote{
Additional boundary counterterms appear when the theory is defined on an
interval. Their significance for our discussion will become clear in the following.}
of dimension $p_i>4$ and $c^{(p_i)}(N_5,\beta)$
is in general a non-perturbative function of $\beta$
and $N_5$ in the interior of the phase diagram.
The sum runs over all independent operators of dimension $p_i$. 
The regularization assumed for this action is one where space--time is
Euclidean and discrete (i.e. $N_5$ is an integer)
and the cut-off is related to the lattice spacing $a$ as $\Lambda=1/a$.
The various operators appearing in $\cal L$ contribute to the masses
of the states that make up the spectrum, in case some fields
acquire a vacuum expectation value (vev). The scalar potential at one-loop
can be computed by inserting the mass matrix into the
Coleman--Weinberg formula.

We expect that there is a region in the parameter space around the
trivial point where the effective lagrangean \eq{efflagran} describes well
the theory, in the sense that a truncation of the expansion is meaningful.
It will turn out to be a good approximation also when we will compare
with results from lattice simulations.

To begin, computing the Coleman--Weinberg potential with a cut-off should
move us somewhere in the interior of the phase diagram. Moreover,
we will seek regions of the phase diagram where a reasonable truncation of the
expansion \eq{efflagran} is possible. The reason is that in such a case, the theory
can be described analytically and non-perturbatively to a good approximation
with a finite number of parameters. In a lattice regularization, 
the leading order value
of the coefficients can be read off by expanding the Wilson plaquette action 
to appropriate order, while the general $c^{(p_i)}(N_5,\beta)$ can be either
computed in perturbation theory or "measured"
by means of a numerical simulation. We will not determine these coefficients
here. Instead we use them as variable parameters of our potential calculation
and compare its results with the lattice data. 

\section{The Coleman--Weinberg potential}

\subsection{Review of the continuum calculation and a few comments}

Let us first remove the cut-off.
The calculation in this case can be carried out in the continuum without 
having to refer explicitly to the lattice regulator.
The Coleman--Weinberg scalar potential $V$
in $D$ dimensions is defined at one-loop by
\bea 
\int [{\rm D}\phi]\; e^{-S_E} & \;\sim\; &
{\rm e}^{-V}\equiv\frac{1}{\sqrt{\det{[\Box + M^2] }}}.\label{pathint}
\eea
This one-loop approximation is guaranteed to be satisfactory
as long as we keep $g_4\ll 1$.
The potential $V$ can be extracted as follows:
\bea
V & = &
-\frac{1}{2}\frac{\pi^{D/2}}{(2\pi)^D}\int_0^\infty
\frac{{\rm d}t}{t^{\frac{D+2}{2}}}{\rm tr}\left\{{\rm e}^{-tM^2}\right\}
\,.\label{CW}
\eea
In order to specify the mass spectrum, one expands the fields
{\`a} la Kaluza--Klein (KK), where 
the eigenvalues of the mass matrix $M$ are of the form
$m_n(\alpha)=(n+f(\alpha))/R$, $n\in\mathbb{Z}$.
In our case the shifts $f(\alpha)$ of the KK numbers are due to some of the 
components of the gauge field $A_5$ taking a vacuum expectation value (vev).

To be more specific, we will consider an $SU(2)$ gauge theory in five
dimensions compactified on an interval of size $\pi R$, which is equivalent
to an $S^1/\mathbb{Z}_2$ orbifold.
The gauge field $A_M=-ig_5A_M^BT^B$ ($T^B=\sigma^B/2$, $B=1,2,3$ are the
$SU(2)$ generators) is defined on the circle $S^1$ along the extra dimension.
Its components $A_M^B$ are divided into even and odd under the orbifold
parity, which is the product of the parity under Euclidean reflection
($+1$ for $A_\mu$, $\mu=0,1,2,3$ and $-1$ for $A_5$) with the parity under
group conjugation ($+1$ for the components with $B=3$ and $-1$ for $B=1,2$).
The conjugation matrix is $g=-i\sigma_3$.
The even $E$ and odd $O$ fields are expanded in the KK basis as
\bea
 E(x,x_5) & = & \frac{1}{\sqrt{2\pi R}}E^{(0)}(x) +
 \frac{1}{\sqrt{\pi R}}\sum_{n=1}^\infty
 E^{(n)}(x)\cos(nx_5/R) \,, \label{evenfields}\\
 O(x,x_5) & = & \frac{1}{\sqrt{\pi R}}\sum_{n=1}^\infty O^{(n)}(x)\sin(nx_5/R) 
 \,. \label{oddfields}
\eea
On the boundary the even fields that survive the orbifold projection
are the scalar (from a four-dimensional point of view)
components $A_5^{1,2}$, which will be our complex Higgs field,
and the gauge boson component $A_\m^3$, which we will call the $Z^0$ gauge
boson.

The mass-squared terms in the lagrangean $\cal{L}$
for the gauge bosons $A_\mu$ are in the continuum 
\bea
 (\bar{D}_5 A^A_\mu)(\bar{D}_5 A^A_\mu) \,,\label{massterm}
\eea
where $\bar{D}_5A_M^A = 
\partial_5A_M^A + g_5 f^{ABC}\langle A_5^B\rangle A_M^C$. 
The masses of the KK modes of $A_\mu$
are obtained by expanding the field components as in \eq{evenfields}
and \eq{oddfields}.           
The mass-squared terms for the scalars $A_5$ originate from the gauge-fixing
term in the lagrangean and are
\bea
 \frac{1}{\xi}(\bar{D}_5 A^A_5)(\bar{D}_5 A^A_5) \,. \label{gf}
\eea
We will work in the gauge $\xi=1$.
In order to find the mass eigenvalues, the gauge symmetry can be used to allow
only one component of $A_5$ to take a vev, here for example
\bea
\langle A_5^1\rangle=\frac{v}{\sqrt{2\pi R}} \,,\quad
\langle A_5^2\rangle=\langle A_5^3\rangle = 0 \,. \label{vevsu2}
\eea
The mass matrix for the four-dimensional fields is diagonal in the KK index
$n$.
The eigenvalue shifts for $n\neq0$ are $f(\alpha)=0,\;\pm\a$ and are the
same for the gauge bosons and the scalars. The parameter $\alpha$
is defined as
\bea
\alpha & = & \frac{g_5 v R}{\sqrt{2\pi R}} \,. \label{alphadef}
\eea
The mass of the "Cartan" component is $m_{Z^0} = \alpha/R$ and the
the masses of the scalar zero-modes are $0,\;\alpha/R$.
The potential \eq{CW} has the simple\footnote{
The derivation of \eq{vacenergy} involves a Poisson resummation of the KK
index. Thereafter a divergent contribution $m=0$ is dropped since it does
not depend on the parameter $\alpha$.}
periodic form \cite{Antoniadis:2001cv,Kubo:2001zc}
\bea
V & = & - \frac{3\cdot P}{64 \pi^6 R^4} \sum _{m=1}^\infty
\frac{\cos{(2\pi m \a)}}{m^5}\,,\label{vacenergy}
\eea
where $P=3$ is the multiplicity of states (2 from physical degrees of
polarization of $A_\mu$, 1 from $A_5$).
Succinctly expressed, the dimensionless modulus in \eq{alphadef}
acquired a non-trivial potential at the quantum level by the Hosotani mechanism
\cite{Hosotani:1983xw,Hosotani:1989bm}.

The fast way to see why the rank can not break at infinite cut-off is to
look at the commutator of the vacuum expectation value of the Polyakov line
around the extra dimension (which is the physical meaning of $A_5$)
with the generators of the Cartan subalgebra:
\be
[e^{-i \pi \alpha \sigma^1} , H_i],
\ee
which clearly vanishes for $\alpha\in\mathbb{Z}$.
From the point of view of the potential,
it is obvious from Eq. (\ref{vacenergy}) that the term that determines the
true vacuum is the first term in the expansion in $m$, proportional to
$\cos{(2\pi \alpha)}$.  
Clearly, \eq{vacenergy} then has a global minimum at $\alpha=\alpha_{\rm min} = 0\; {\rm mod}\, \mathbb{Z}$ as a
result of which the KK tower shifts by an integer and thus
SSB of the rank of the gauge group does not occur, consistently with the symmetry argument.

Nevertheless, a non-trivial Higgs mass is implied by $V$. 
In fact, the Higgs mass formula found by computing the second derivative 
of the potential \eq{vacenergy} at the minimum agrees with the one obtained
by a direct continuum perturbative calculation of vacuum polarization diagrams
in dimensional regularization \cite{vonGersdorff:2002as,Cheng:2002iz}
\be
(m_H R)^2_{\rm pert} \;\equiv\;
\left.R^2\frac{{\rm d}^2V}{{\rm d}v^2}\right|_{\alpha=\alpha_{\rm min}} \;=\;
\left.\frac{9 \zeta(3)}{8 \pi^4}\frac{1}{N_5\beta}\right|_{\alpha=0}
\label{su2mhpert}
\ee
expressed in terms of the dimensionless parameters.\footnote{Note 
that the squared Higgs mass on the orbifold is just one half of that
on the circle \cite{Puchwein:2003jq}.}
We are particularly interested in the ratio
\bea
\r_{HZ^0} & = & \frac{m_H}{m_{Z^0}} \,,
\eea
whose one-loop value is
naturally undetermined at the trivial point.

To summarize, at the trivial point, the $Z^0$ gauge boson remains massless and
the Higgs is inclined towards masslessness as well. These phenomenological 
obstacles are expected to persist generically as long as
the theory is probed in the close vicinity of the trivial point.

It is important therefore to investigate if these are properties
valid throughout the entire phase diagram. For instance, in order that the rank is
preserved non-perturbatively, a discrete global symmetry must act on $\alpha$
so that it can protect the perturbative minimum of the potential. We see no
such symmetry. In fact, a careful lattice simulation study of an $SU(2)$
orbifold gauge theory, for a range of finite values of $N_5$ and $\beta$ shows
that there is no leftover massless $U(1)$ gauge boson, suggesting a 
non-perturbative breaking of the rank \cite{Irges:2006zf,Irges:2006hg}.  
As a simple plausibility argument, let us assume that 
$\alpha$, not being protected by any symmetry, changes
from its integer value at the trivial point by a cut-off dependent shift
$\epsilon (\L)$ in the interior of the phase diagram. Then, 
\bea
[e^{-i \pi (1-\e(\L)) \sigma^1} , H_i] & \sim & {\cal O}(\e(\L)),
\eea
the Polyakov loop does not commute anymore with the $H_i$, allowing for
SSB of the rank, which can occur even if $\epsilon (\L)$ is infinitesimal.

As we will show in the remaining of this letter, the transition 
from the Coulomb into the Higgs phase on the circle is sharp
with the modulus $\a$ shifting discretely due to the presence
of cut-off dependent higher dimensional operators. 
On the orbifold, the minimum is further shifted continuously due to boundary counterterms,
which also bring important cut-off effects.

\subsection{Cut-off effects in the mass formulae}

As a first step, we would like to derive the first non-trivial 
correction in the effective action which determines the
leading correction to the mass formula. For simplicity we consider the
theory regulated on an infinite lattice. The $SU(N)$ Wilson plaquette action is
\bea
S_{\rm W} & = & -\frac{\beta}{2N} \sum_z \sum_{M,N}
{\rm tr}\{2-U_{MN}(z)-U^\dagger_{MN}(z)\}
\,,
\eea  
where $z=(x_\mu,x_5)$ and $\mu=1,\ldots,4$, $M=1,\ldots,5$.
The plaquettes are defined as
$U_{MN}(z)=U(z,M) U(z+a{\hat M},N) U^{-1}(z+a{\hat N},M) U^{-1}(z,N)$ and 
a link is related to the gauge field via $U(z,M)=e^{a\, A_M(z)}$.
Using the Baker--Campbell--Hausdorff formula, given the
Lie algebra elements $X,Y,P,Q$ one can determine the $F_i$ such that
\bea
e^{aX}e^{aY}e^{aP}e^{aQ} & = & e^{\sum_{i} a^i F_i} \,,
\eea
where $i$ runs from 1 to some specified order $I$, with all the $F_i$ 
anti-hermitian. This implies that in expanding the plaquette action with
$X=A_M(z)$, $Y=A_N(z+a{\hat M})$, $P=-A_M(z+a{\hat N})$ and $Q=-A_N(z)$,
the non-zero contributions arrange themselves as
\bea
S_{\rm W} & = & \frac{\beta}{2N} \sum_{n=1}^{\infty}\frac{2}{(2n)!}
\left(\sum_{i=1}^{I} a^i F_i\right)^{2n} \,.
\eea
Let us start from the $n=1$ term. This is $(aF_1 + a^2 F_2 + \ldots)^2$,
whose leading order part, by substituting the explicit forms
\be
F_1= A_M(z)+A_N(z+a{\hat M})-A_M(z+a{\hat N})-A_N(z),\hskip .5cm 
F_2=[A_M(z),A_N(z)] \,,
\ee
is recognized to correspond to the dimension 4 operator 
${\cal O}^{(4)}= \sum_{M,N}{\rm tr}\{F_{MN}F_{MN}\}$.
The next terms in the expansion generate the dimension 6 operator
\cite{Lepage:1996jw}
\bea 
{\cal O}^{(6)} & = &
\sum_{M,N} \frac{1}{24}{\rm tr}\{F_{MN}(D_M^2+D_N^2)F_{MN}\} \label{op}
\eea
and so forth. The squared gauge boson masses (when $A_\mu$ is expanded
in the KK basis) can be extracted by diagonalizing the operator
\be
\bar{D}_5\bar{D}_5 + \frac{a^2}{12} (\bar{D}_5\bar{D}_5)^2 +\ldots \,, 
\label{opm}
\ee
where $\bar{D}_5 = \partial_5  + [\langle A_5\rangle, \; \cdot\; ]$.
The mass squared matrix is therefore itself an expansion
\bea
M^2 & = & (M^2)^{(4)} + \frac{a^2}{12}(M^2)^{(6)} + \ldots \,, \quad 
(M^2)^{(6)}=\left((M^2)^{(4)}\right)^2
\eea
with the superscript denoting the dimension of the contributing operator.  
As mentioned, non-perturbatively, the coefficient $1/12$ in \eq{opm}
should be replaced by a generic coefficient
\be
c\equiv c^{(6)}(N_5,\beta) \,.
\ee
The masses of the field $A_5$ come entirely from the gauge fixing term and
do not receive corrections from the action.

On the orbifold, an additional contribution to the mass matrix comes from the
boundary counterterm
\bea
{\cal L}^{{\rm bound.}} & = & \frac{\pi a{c}_0}{4}
F_{5\mu}^{\hat a}F_{5\mu}^{\hat a} \left[\delta(x_5) + \delta(x_5-\pi R)\right]
\label{boundaryterm}
\eea
in the effective lagrangean \eq{efflagran}.
It will give a contribution only to
the mass matrix of the even gauge fields $A_\mu^a$ because these do not vanish
on the boundaries. We note that the counterterm \eq{boundaryterm} does not
appear at one-loop in perturbation theory
\cite{vonGersdorff:2002as,Cheng:2002iz}
but is expected to arise at higher orders \cite{Irges:2004gy}.
The boundary coefficient should also be understood as 
a cut-off dependent function ${c}_0(N_5,\b)$ at a generic point of the phase diagram.

It is straightforward to obtain the cut-off corrected eigenvalues of the mass matrices.
We will consider two models. One is the $SU(2)$ model we have already described and the other
is its $SU(3)$ generalization. In this case the orbifold breaks the symmetry down to $SU(2)\times U(1)$ on the
boundaries. The even fields are $A_\m^{1,2,3,8}$ and $A_5^{4,5,6,7}$, the rest of them are odd. 
The vev can be aligned along
$\langle A_5^4\rangle=v/\sqrt{2\pi R}$ and the eigenvalues are
again parameterized by $\alpha$ defined in \eq{alphadef}.

For the $SU(2)\stackrel{\mathbb{Z}_2}{\rightarrow}U(1)$ model the non-zero mode 
eigenvalues of the mass matrix $(MR)^2$ are (we recall the relation 
$\pi R=N_5a$)
\bea
(m_nR)^2 =
&& n^2\, ,\\
&& (n\pm \alpha)^2 + \frac{c_0\a^2}{2}\frac{\pi}{N_5}+
c\, (n\pm \alpha)^4\frac{\pi^2}{N_5^2} \, . \label{bosonn}
\eea
The single zero mode eigenvalue $(m_{Z^0}R)^2$ can be obtained by putting $n=0$ in \eq{bosonn}.
For the $SU(3)\stackrel{\mathbb{Z}_2}{\rightarrow}SU(2)\times U(1)$ model 
in the non-zero mode sector we find  ($2\times$ denotes degeneracies)
\bea
(m_nR)^2 =
 &&2\times n^2\,,\\
 &&(n\pm\alpha)^2+\frac{c_0\alpha^2}{4}\frac{\pi}{N_5}+c\,(n\pm \a)^4\frac{\pi^2}{N_5^2}\,,\label{Z}\\
 &&2\times \left( (n\pm\alpha/2)^2+\frac{c_0(\alpha/2)^2}{4}
 \frac{\pi}{N_5}+c\, (n\pm {\a}/{2})^4\frac{\pi^2}{N_5^2}\right)\,.\label{W}
\eea
In the zero-mode sector there is a single zero eigenvalue (corresponding to $(m_\g R)^2$), one eigenvalue 
which is obtained by putting $n=0$ in Eq. (\ref{Z}) 
(corresponding to $(m_{Z^0}R)^2$) and a pair of eigenvalues which is obtained
by putting $n=0$ in Eq. (\ref{W}) (corresponding to $(m_{W^\pm}R)^2$).
The only subtle step in these calculations is that for the boundary contributions we have kept only their
$n$-independent parts and dropped O($1/n$) corrections,
which seems to be a satisfactory approximation.

Another new input immediately appears because of the finite cut-off.
Namely, the Higgs vev should not exceed $1/a$. This translates in the constraint
\be
|\alpha| < \sqrt{\frac{N N_5}{\pi^2 \b}} \label{constraint}
\ee  
and the potential we are about to compute strictly makes sense only in this
regime.

\subsection{Coleman--Weinberg potential with a cut-off}

We consider a one-component scalar field of mass $m$ on a $D$ dimensional
Euclidean lattice with lattice spacing $a$. The effective potential can be
written in the same form as in the continuum \eq{CW}
\bea
\label{pot0}
V & = & - \frac{1}{2} \int \limits_\epsilon^\infty \frac{\dt}{t} 
{\rm tr}\left[{\rm e}^{-t(m^2 + \hat{p}^2)}\right],
\eea
but here $\hat{p}_\mu = \frac{2}{a} \sin{(\frac{a p_\mu}{2})}$ are the
lattice momenta and the $p_\mu$ take values restricted to the interval
$-\pi/a$ and $\pi/a$. Hence \eq{pot0} is short for
\bea
\label{v1}
V & = &  - \frac{1}{2} \int \limits_0^\infty \frac{\dl}{l} 
{\rm e}^{-\frac{1}{l}(m^2a^2 + 2D)}\frac{1}{a^D}{\rm I}_0^D\left(\frac{2}{l}\right)
\,,
\eea
where we have substituted $1/t$ by $l/a^2$ and ${\rm I}_0(2/l)$ is the $0^{\rm
  th}$
modified Bessel function.
The total potential $V$ is the sum over all mass states of the ghosts, gauge and scalar particles.

\subsubsection{The $SU(2)$ case}

Let us start with the gauge bosons $A_\mu$.
We include only the $n$ dependent terms of the mass eigenvalues of \eq{bosonn}
into $m_n$. This leads to
the modified KK masses
\be
\label{kkmass}
 m_n^2 = \left(\frac{n+\alpha}{R}\right)^2 + c \ a^2
 \left(\frac{n+\alpha}{R}\right)^4.
 \ee
The boundary correction is accounted for by a shift in the
exponential in Eq.~(\ref{v1}), i.e. (setting $D = 4$)
 \be
 e^{-\frac{1}{l}(m_n^2a^2 + 8)} \to e^{-\frac{1}{l}(m_n^2a^2 + 8 + c_0 \frac{a^3
     \alpha^2}{2 R^3})}.
 \ee
The $c_0$ term is of $O(a)$ relative to the mass squared 
and is therefore expected to be small, such that it can be expanded.
At the same time, we use an expansion in $c$ 
throwing away
all terms of $O(a^3)$ or higher in the joint expansion. 
Schematically, we will collect the terms
\be
\label{Vexp}
V  = \underbrace{f_0}_{O(1)} + \underbrace{f_2 \ c }_{O(a^2)} 
+ \underbrace{f_1 \ c_0 }_{O(a)} + \underbrace{\tilde{f}_2 \ c_0^2}_{O(a^2)}.
\ee
We start with the pure bulk contributions, that is $f_0$ and $f_2$.
On the lattice the values of $n$ are restricted to the interval $n = 0,1,..,
N_5-1$ but we extend them to $n \to \infty$ for the calculation. This is
justified because the higher modes are expected to 
contribute very little \cite{Dienes:1998vg}.
Hence, summing over all KK states gives
\be
\label{vorig}
V^{\rm bulk} =  -\frac{1}{2a^4}\sum \limits_{n \in \mathbb{Z}}\int \limits_0^\infty \frac{\dl}{l} e^{-\frac{1}{l}(m_n^2a^2 +
  8 )}I_0^4\left(\frac{2}{l}\right).
\ee 
Note that on the orbifold the extension to the sum over $n \in \mathbb{Z}$ is
achieved by combining the $+ \alpha$ and $-\alpha$ contributions in the
eigenvalues \eq{bosonn}.
We expand in the $O(a^2)$ correction to the mass formula \eq{kkmass} and obtain
\be
V^{\rm bulk} = - \frac{1}{2a^4} \sum\limits_{k= 0}^{\infty} c^k \frac{(-a^4)^{k}}{k! R^{4k}}
\int \limits_0^\infty \frac{\dl}{l} \frac{1}{l^k} e^{-\frac{8}{l}}
\I\left(\frac{2}{l}\right) \sum \limits_{r =  0}^{4k} {4k \choose r} \alpha^{4k-r}
\sum\limits_{n \in\mathbb{Z}} e^{-\left(\frac{n+\alpha}{R}\right)^2\frac{a^2}{l}} 
n^r, \label{Vbulk}
\ee
where we have expressed $(n + \alpha)^{4k}$ through the binomial identity.
The next step is a Poisson resummation.
From the ordinary Poisson resummation formula follows the identity of its derivatives
($A$ and $b$ are constants and $m$ an integer not to be confused with the mass $m$)
\be
\label{poi_deriv}
\frac{1}{\pi^r} \frac{\partial^r}{\partial b^r} \sum_n e^{-(\pi A n^2 + \pi b
  n)} = \frac{1}{\pi^r} \frac{\partial^r}{\partial b^r} \sqrt{A}^{-1} \sum_m
e^{-\frac{\pi}{A}(m + ib/2)^2}.
\ee
The $\sum_n$ appears in \eq{poi_deriv} exactly in the same way as in the rhs of \eq{Vbulk},
hence we need to find the resummation of the exponential alone and then take
its $r$-th derivative.
We define functions $f^{(r, m)}(\alpha, l)$ by the relation
\be
f^{(r, m)}(\alpha, l): \  \frac{\partial^r}{\partial b^r}
e^{-\frac{\pi}{A}(m+i\frac{b}{2})^2} = f^{(r, m)}(\alpha, l) e^{-\frac{\pi}{A}(m+i\frac{b}{2})^2}.
\ee
These functions have the property that $f^{(r, -m)}(\alpha, m) = [f^{(r,
  m)}(\alpha, m)]^*$. Thus $V^{\rm bulk}$ can be written as
\be
\label{pot1}
V^{\rm bulk} = -\frac{1}{2} \frac{R}{a}\sum_k (-1)^k c^k V_k,
\ee
where 
\be
\begin{split}
V_k =& \frac{a^{4(k-1)}}{k! R^{4k}}\sum \limits_{r = 0}^{4k} 
\frac{1}{\pi^{r-\half}}{4k \choose r} \alpha^{4k-r} \int_0^\infty \dl l^{-\half - k}
e^{-\frac{8}{l}}
I_0^4\left(\frac{2}{l}\right) \times\\
&\times \left(f^{(r, 0)}(\alpha, l) + \sum\limits_{m > 0} 
2 \, {\rm Re} \left[ f^{(r,m)}(\alpha, l) e^{-2i\pi m\alpha}\right]e^{-\frac{\pi^2 R^2 lm^2}{a^2}}\right).
\end{split}
\ee
To be consistent in the order of the expansion we must truncate this sum at $k=1$.
A fact that one should notice here and keep in mind is that the 
$k=1$ term comes with the opposite sign with respect to the $k=0$ term.

The original integral in \eq{v1} has a logarithmic divergence for $l
\to \infty$. 
The Poisson resummation regularizes the integral in
the sense that after the resummation
it is finite up to a divergent term coming from $m = 0$. This term is
constant in $\alpha$ and can be neglected. 
Hence the terms corresponding to $f_0$ and $f_2$ are 
\be
\label{v0_nc}
{V}_0 =  \frac{2 \sqrt{\pi}}{a^4} \int_0^\infty \frac{\dl}{\sqrt{l}} \  \ e^{-\frac{8}{l}}\
\I\left(\frac{2}{l}\right) \ \sum \limits_{m>0}  \cos{(2\pi  m
  \alpha)}\ e^{-N_5^2 m^2 l}
\ee
and
 \be
\label{v1_nc}
V_1 = \frac{2 \sqrt{\pi}}{a^4}
\int \dl \sqrt{l }\  e^{-\frac{8}{l}} \ \I
\left(\frac{2}{l}\right) \sum \limits_{m>0} 
  \cos{(2 \pi m \alpha)} \left(N_5^4 m^4 l^2 -3N_5^2 m^2 l
    +\frac{3}{4}\right)e^{-N_5^2 m^2 l}
\ee
respectively, where we used $R \pi = N_5 a$. 
For the boundary potential $V^{\rm bound.}$ determined by the terms $f_1$ and 
$\tilde{f}_2$ 
we need the first and second order terms in $c_0$ at $c = 0$.
After a Poisson resummation, they look very similar to $V_0$ in \eq{v0_nc}.
Apart from a prefactor, 
the only difference is the modified power of $l$ in the integrand.

Summing up, we finally obtain $V^{\rm gauge}=V^{\rm bulk}+V^{\rm bound.}$,
or explicitly
\be
\label{vgauge}
\begin{split}
V^{\rm gauge} &= -\half \frac{N_5}{\pi} \left({V}_0 - c {V}_1\right) + \\
&+ \frac{c_0}{4} \frac{\a^2 \pi^{5/2}}{N_5^2} \int_0^\infty \dl \ l^{-3/2} e^{-\frac{8}{l}}\I
  \left(\frac{2}{l}\right)\left(1+2\sum
    \limits_{m>0}\cos{(2\pi m\alpha)}e^{-{N_5^2 lm^2}}\right)\\
& - \frac{c_0^2}{16}\frac{\a^4 \pi^{11/2}}{N_5^5} \int_0^\infty \dl \ l^{-5/2} e^{-\frac{8}{l}}\I
  \left(\frac{2}{l}\right)\left(1+2\sum
    \limits_{m>0}\cos{(2\pi m\alpha)}e^{-{N_5^2 lm^2}}\right)\ 
\end{split}
\ee
per gauge field $A_\mu$.

Note that in the boundary corrections the $m=0$ term can not be dropped because
it is $\alpha$-dependent.
However, the higher negative power in $l$ makes it converge as $l\rightarrow \infty$.
On the other hand, the $c_0^2$ correction in \eq{vgauge} diverges in the limit
$l \to 0$. 
This can be dealt with by
splitting the corresponding integral in two domains.
For small $l$ one can use \eq{vorig} directly, with $m_n^2$ set to
$(n+\alpha)^2/R^2$
in which case the only potential problems arise when $n+\alpha = 0$.
This however is not a real problem
because on one hand at $\alpha = 0$ the $c_0$ corrections vanish 
identically and 
on the other $\alpha$ will be always smaller than one due to the
constraint in \eq{constraint}. For large $l$ one can then use \eq{vgauge}.

The mass matrices for $A_5$ and the Faddeev--Popov ghosts are
identical\footnote{
Only the orbifold parities are opposite. Those of the ghosts are equal to the
ones of the gauge bosons, as it follows by considering their interaction term
in the lagrangean.}
and generated through $\bar{D}_5^2$ only. Moreover they do not get cut-off
corrections from the bulk and boundary action.
The contribution to the potential for $A_5$ is
\bea
V^{\rm scalar} & = & V^{\rm gauge}|_{c_0=c=0} \,.
\label{vscalar}
\eea
In the notation of \eq{pathint} the 1-loop path integral for our theory is
\be
 \frac{\det[\prod_n (-\partial_\mu\partial_\mu + m_{n,{\rm scalar}}^{2})]  }
{\det[\prod_n (-\partial_\mu\partial_\mu + m_{n,{\rm gauge}}^2)]^{D/2}  
\det[\prod_n(-\partial_\mu\partial_\mu + m_{n,{\rm scalar}}^2)]^{1/2}  
}
\ee
where $m_{n,{\rm gauge}}$ is given by \eq{bosonn} and
$m_{n,{\rm scalar}}=m_{n,{\rm gauge}}|_{c_0=c=0}$.
For $D = 4$ this is equivalent to a Coleman--Weinberg potential
\be
V  = 4 V^{\rm gauge} + V^{\rm scalar} - 2 V^{\rm scalar}\label{Vtotal},
\ee
where the last (negative) contribution comes from the ghosts.

\subsubsection{The $SU(3)$ case}

The eigenvalues of the $SU(3)$ mass matrix \eq{Z} and \eq{W} 
are very similar to those of the $SU(2)$ model in \eq{bosonn}.
They can be obtained 
by shifting $c_0 \to c_0/2$ and for $W$ in addition 
by setting $\alpha \to \alpha/2$.
Otherwise the derivation of the potential is exactly the same as in
the $SU(2)$ case. 
We obtain for the $Z^0$
\be
\begin{split}
V_{Z^0}^{\rm gauge} &= -\half\frac{N_5}{\pi}\left(V_0 - c V_1\right)\\
&+\frac{c_0}{8} \frac{\alpha^2 \pi^{5/2}}{a^4 N_5^2}\int \dl l^{-3/2}e^{-8/l}
\I\left(\frac{2}{l}\right)
\left(1 + 2 \sum \limits_{m>0} \cos{(2\pi m \alpha)} e^{-N_5^2 m^2 l}\right)\\
&- \frac{c_0^2}{64} \frac{\alpha^4 \pi^{11/2}}{a^4 N_5^5} \int \dl l^{-5/2} e^{-8/l}
\I\left(\frac{2}{l}\right)
\left(1 + 2 \sum \limits_{m>0} \cos{(2 \pi m \alpha)} e^{-N_5^2 m^2 l}\right)
\end{split}
\ee
with $V_0$ and $V_1$ as in \eqs{v0_nc}{v1_nc}.
Scalar and ghost contributions are exactly the same as for $SU(2)$ because
they do not see the $c_0$ corrections anyway. 
In the case of the $W$ bosons the two different parts of the gauge bulk contributions are
\be
V_0^{W^\pm} = \frac{2 \sqrt{\pi}}{a^4} \int \frac{\dl}{\sqrt{l}} e^{-8/l} \I \left(\frac{2}{l}\right)
\sum \limits_{m>0}\cos{(\pi m \alpha)} \, e^{-N_5^2 m^2 l}
\ee
and
\be
V_1^{W^\pm} = \frac{2 \sqrt{\pi}}{a^4}\int \dl \, \sqrt{l}\, e^{-8/l} \I
\left(\frac{2}{l}\right) \sum \limits_{m>0} \cos{(\pi m\alpha)} \left(N_5^4 m^4
  l^2 - 3 N_5^2 m^2 l + \frac{3}{4}\right) e^{-N_5^2 m^2 l}.
\ee
Finally, adding the boundary corrections leads to
\be
\begin{split}
V_{W^\pm}^{\rm gauge} &= -\half\frac{N_5}{\pi}\left[{V}_0^{W^\pm} - c {V}_1^{W^\pm}\right]\\
&+\frac{c_0}{32} \frac{\alpha^2 \pi^{5/2}}{a^4 N_5^2}\int \dl l^{-3/2} e^{-8/l}
\I\left(\frac{2}{l}\right)
\left(1 + 2 \sum \limits_{m>0} \cos{(\pi m \alpha)} \, e^{-N_5^2 m^2 l}\right)\\
&- \frac{c_0^2}{1024} \frac{\alpha^4 \pi^{11/2}}{a^4 N_5^5} \int \dl l^{-5/2} e^{-8/l}
\I\left(\frac{2}{l}\right)
\left(1 + 2 \sum \limits_{m>0} \cos{( \pi m \alpha)} \, e^{-N_5^2 m^2 l}\right).
\end{split}
\ee
Naturally, here the scalar/ghost contributions are also modified accordingly,
that is we have to replace  $\cos{(2 \pi m \alpha)}$ by $\cos{(\pi m \alpha)}$
in \eq{vscalar}.
Summing up, the total Coleman--Weinberg potential for $SU(3)$ is
\be
V = 4 V_{Z^0}^{\rm gauge} + 8 V_{W^\pm}^{\rm gauge} 
    - 2 V_{W^\pm}^{\rm scalar} - V^{\rm scalar}.
\ee
The mode counting includes an additional factor of 2 for the degenerate  $W$
bosons. 

\section{Numerical results}
\begin{figure}[!t]
\begin{center}
\includegraphics[width=4in]{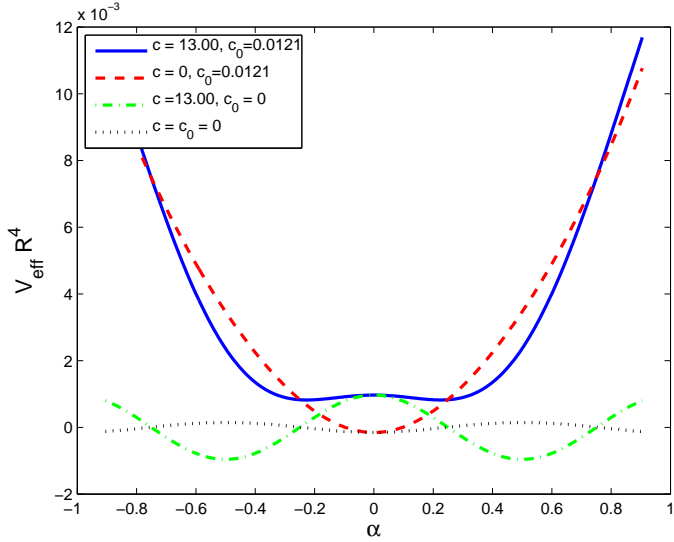}
\end{center}
\caption{\small The Coleman--Weinberg potential for $SU(2)$ at $N_5=6$. 
 The plot shows four different cases: the potential with and without boundary
 and bulk corrections. The parameters are $c=13.0$ and
 $c_0=0.0121$, if not set to zero. The minimum for the full potential
 (solid line) lies at $\alpha_{\rm min}=0.225$.}
\label{f_potential}
\end{figure}

\fig{f_potential} shows the Coleman--Weinberg potential for $SU(2)$,
\eq{Vtotal}, 
choosing $N_5=6$, $c=13.0$ and $c_0=0.0121$. We plot
the dimensionless product $R^4 V$.
From the structure of \eq{vgauge} we can see that on the circle (where $c_0=0$)
for large enough $c$, the term $c V_1$ dominates
over $V_0$ and there is a discrete shift of the minimum from $\alpha=0$ to 
$\alpha=1/2$. 
For $N_5=6$, if $c$ is set to its tree-level value $1/12$ in the
Wilson plaquette effective action, there is no SSB.
The critical value is at $c \approx 1.72$. Given the
non-renormalizability of the theory, there are quantum corrections which are
power-like in the cut-off and large discrepancies from the tree level value are
perhaps not unexpected.
That is on $S^1$, the transition from the Coulomb into the Higgs phase is sharp
and the minimum shifts by half an integer.
On $S^1/\mathbb{Z}_2$ (where $c_0 \ne 0$) the non-trivial minimum can drift
continuously away from $\alpha=1/2$ due to the $\alpha$-dependent prefactors
in $V^{\rm bound.}$.
In general, for a given
$c$ there is a maximal value of $c_0$ for which there is SSB, that is the
potential has a minimum at $0<\alpha_{\rm min}<1$. For $c_0$ larger than
this value the minimum is at $\alpha=0$.
These expectations are confirmed by \fig{f_potential},
which also shows that it is not possible to have SSB due to boundary
effects only, that is with $c_0>0$ and $c=0$.

\begin{figure}[!t]
\begin{center}
\includegraphics[width=4in]{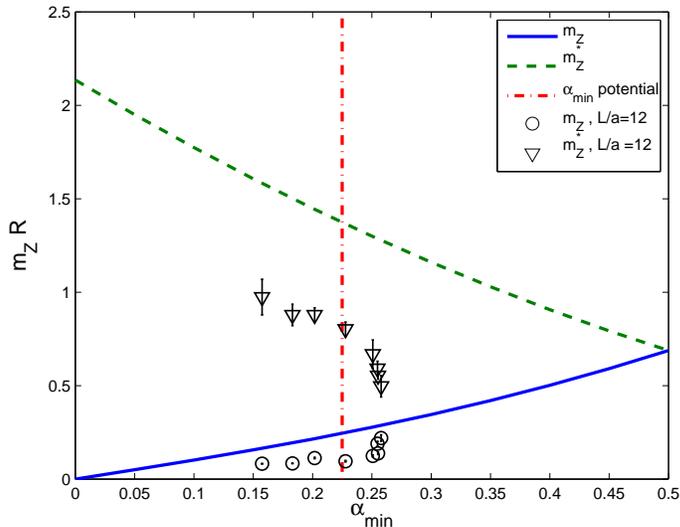}
\end{center}
\caption{\small
$m_Z$ and $m_Z^*$ as function of $\alpha_{\rm min}$ for $SU(2)$ at $N_5=6$.
Comparison of the analytical formula \eq{mZ} with the lattice data at
$L/a=12$ and $T/a=96$.}
\label{f_mz_alpha}
\end{figure}

The masses of the $Z^0$ boson and its first excited state $Z^{0*}$ correspond
to the two lightest modes in \eq{bosonn} in the $SU(2)$ model. They depend on
the value of $\alpha_{\rm min}$ which minimizes the potential in \eq{Vtotal}.
We have
\be
 m_{Z^0} R = \min_{n \in \{0,1\}}
 \sqrt{(n-\alpha_{\rm min})^2 + \frac{c_0 \pi}{2N_5}\alpha_{\rm min}^2 +
 \frac{c \pi^2}{N_5^2}(n-\alpha_{\rm min})^4}
 \label{mZ}
\ee
and the mass of the first excited state $m_{Z^{0*}}$ corresponds to the 
$\max \limits_{n    \in\{0,1\}} $ in the above formula.
In \fig{f_mz_alpha} we compare the analytical formulae \eq{mZ} for
$N_5=6$, $c=13.0$ and $c_0=0.0121$ (solid lines) with the lattice results. 
We denote the number of points in the lattice along the spatial directions
by $L/a$, along the time direction by $T/a$ and along the extra dimension
by $N_5$ with $N_5\le L/a$. 
We always compare results from the analytical formulae
and the lattice with the same $N_5$. Increasing $L/a$ with $N_5$ fixed 
corresponds then to compactifying the extra dimension on the lattice.

\begin{figure}[!t]
\begin{center}
\includegraphics[width=4in]{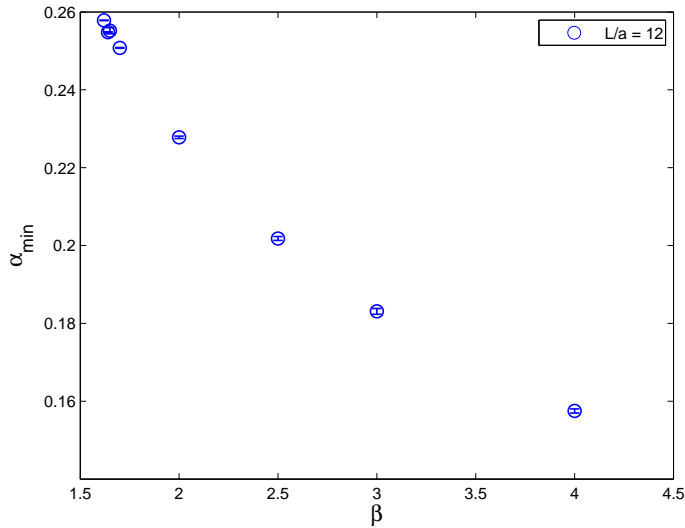}
\end{center}
\caption{\small 
The function $\alpha_{\rm min}(\beta)$ for $SU(2)$ at $N_5=6$ determined from
\eq{alphalat} and lattice simulations with $L/a=12$ and $T/a=96$.}
\label{f_alphabeta}
\end{figure}

The black symbols are our numerical data for $N_5=6$, $L/a=12$ and $T/a=96$.
The points at $\alpha_{\rm min}\approx0.25$ correspond to small $\beta$ values 
near the phase transition.
On the lattice we define the Higgs field $\Phi$ by
\be
\Phi = [a A_5, g] \,,\quad a A_5=\frac{1}{4N_5}(P-P^\dagger)\,,
\ee
where $g$ is the orbifold projection matrix and $P$ the
Polyakov line\footnote{
We take the product of the gauge links along the extra dimension
with no displacement and no smearing.
For details on the lattice construction and the measurement
of its observables, see \cite{Irges:2006hg}.}.
We measure
$\langle \tr{\{\Phi \Phi^\dag\}} \rangle = 2 a^2 g_5^2\langle A_5^1\rangle^2$.
From \eq{vevsu2} and \eq{alphadef} it follows that this observable
can be expressed through $\alpha_{\rm min}$ as
\be
\langle \tr{\{\Phi \Phi^\dag\}} \rangle = 2 \alpha_{\rm min}^2 \frac{\pi}{N_5^2} \,. 
\label{alphalat}
\ee
If we take the values of $\langle \tr{\{\Phi \Phi^\dag\}} \rangle$
from lattice simulations,
\eq{alphalat} can be understood as an implicit function
$\alpha_{\rm min}(\beta)$ for a given orbifold geometry.
For $N_5=6$, $L/a=12$ and $T/a=96$ we plot this function in \fig{f_alphabeta}.
We can then take the $Z$ and $Z^*$ masses from the same simulations
and plot them in \fig{f_mz_alpha}.
The agreement with the analytical formula \eq{mZ} is not so bad.

The Coleman--Weinberg potential has a minimum at
$\alpha_{\rm min} =0.225$ for the chosen $c$ and $c_0$. This value is marked
by the dash-dotted vertical line in \fig{f_mz_alpha}.
In our comparison we are taking the ``bare'' values of
$\alpha_{\rm min}$ to be the same in the potential and in the lattice
computations. There might be a finite renormalization factor relating them.
\begin{figure}[!t]
\begin{center}
\includegraphics[width=4in]{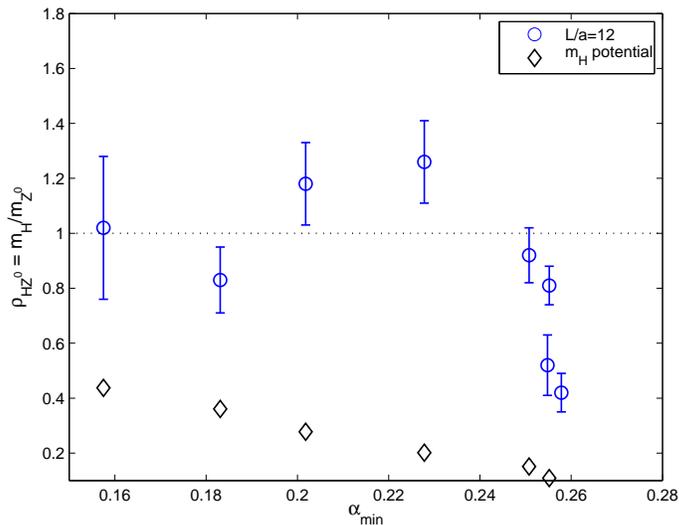}
\end{center}
\caption{\small 
Ratio of the Higgs and the $Z^0$ boson masses for $SU(2)$ at $N_5=6$.
Comparison of the lattice data with the Coleman--Weinberg potential
computation.}
\label{f_ratio}
\end{figure}

From the Coleman--Weinberg potential we can derive the Higgs mass by taking
the second derivative, cf. \eq{su2mhpert}. For a general gauge group $SU(N)$
it is
\be
 (m_HR)^2 =
 \frac{N}{N_5\beta}R^4
 \left.\frac{{\rm d}^2V}{{\rm d}\alpha^2}\right|_{\alpha=\alpha_{\rm min}}\,,
 \label{Higgs_CW}
\ee
where we used $g_5^2/(2\pi R)=N/(N_5\beta)$.
The second derivative is evaluated numerically at the minimum
$\alpha=\alpha_{\rm min}$. Using the $Z^0$ mass from \eq{mZ} we obtain values
for the ratio $\rho_{HZ^0}=m_H/m_{Z^0}$. In \fig{f_ratio} we compare these
values with the lattice data at $N_5=6$. The parameters $c$ and $c_0$ in the
Coleman--Weinberg potential are tuned such that the position of the minimum 
$\alpha_{\rm min}$ matches the value of the lattice data for $L/a=12$, see
\fig{f_alphabeta}, and the corresponding value of $\beta$ is inserted in
\eq{Higgs_CW}. The ratio $\r_{HZ^0}$ computed
in the lattice simulations is significantly larger than the one computed
with the Coleman--Weinberg potential. This is to be attributed to the values
of the Higgs mass extracted from the potential, which are too small.
On the lattice it is possible to get values of $\rho_{HZ^0}\ge1$.
\begin{figure}[!t]
\begin{center}
\includegraphics[width=4in, height=3in]{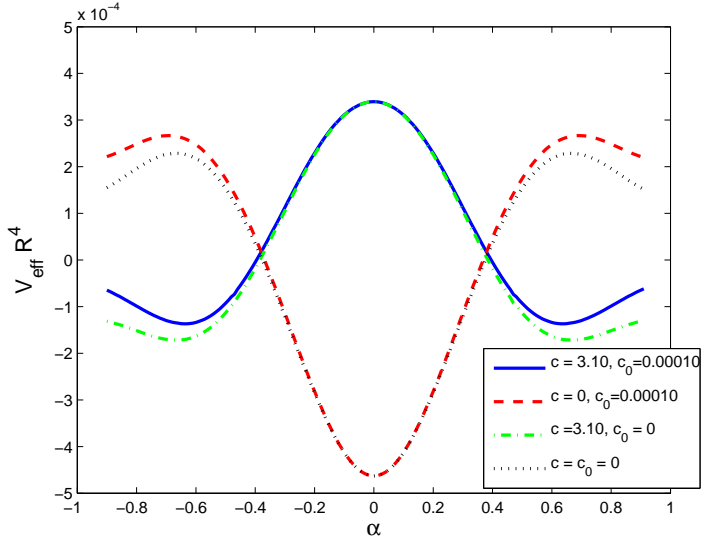}
\end{center}
\caption{\small The Coleman--Weinberg potential for $SU(3)$ at $N_5=6$.
The minimum for $c=3.1$ and $c_0=0.0001$ is at $\alpha_{\rm min} = 0.64$. }
\label{f_potential_SU3}
\end{figure}

Finally, we consider the $SU(3)$ case. 
Here since we do not have lattice data yet, we will 
try to fit the analytical predictions to the experimental data.
In \fig{f_potential_SU3} we
plot the Coleman--Weinberg potential again in the dimensionless form $V R^4$
for $N_5=6$, $c=3.10$ and $c_0=0.0001$.
The masses of the $Z^0$ and $W^\pm$ bosons are given by
\bea
 && m_{Z^0}R = \min_{n\in\{0,1\}}\sqrt{(n-\alpha_{\rm min})^2 +
 \frac{c_0\pi}{4N_5}\alpha_{\rm min}^2+ 
 \frac{c\pi^2}{N_5^2}(n- \alpha_{\rm min})^4} \label{mZSU3} \, \\
 && m_WR = \min_{n\in\{0,1\}}\sqrt{
 \left(n-\frac{\alpha_{\rm min}}{2}\right)^2 +
 \frac{c_0\pi}{4N_5}\left(\frac{\alpha_{\rm min}}{2}\right)^2+ 
 \frac{c\pi^2}{N_5^2}\left(n-\frac{\alpha_{\rm min}}{2}\right)^4}
 \nonumber\\
 && \label{mWSU3}
\eea
(the choice of $n$ that maximizes the expressions gives the mass of the
excited states).
It is easy to see that
for $\alpha_{\rm min}>0.5$ the $Z^0$ mass corresponds to $n=1$ in \eq{mZSU3}.
The $W$ mass is always given by $n=0$ in \eq{mWSU3}.
Therefore the cosine of the Weinberg angle
\bea
 \cos(\theta_{\rm W}) & = & \frac{m_W}{m_{Z^0}} \label{Wein}
\eea
gets larger than the value $1/2$ which is the predicted value from the mass eigenvalues for
$c=c_0=0$. The experimental value of the cosine of the Weinberg angle is
$\cos(\theta_{\rm W}^{\rm exp.})\simeq 0.877$ and can in principle be reached 
if the minimization of the potential gives the appropriate value of
$\alpha_{\rm min}$, which turns out to be close to 0.65.
Our general observation for $N_5=4,\;6$ or higher is that if we choose $c$
and $c_0$ such that the Weinberg angle is close to the experimental value, the
Higgs mass\footnote{
At one-loop order the Coleman--Weinberg potential does not depend on the
gauge coupling but the Higgs mass, as defined in \eq{Higgs_CW}, does.
For the $SU(3)$ pure gauge theory in five dimensions, the phase transition 
with periodic boundary conditions is at $\beta_c=4.35(15)$
\cite{Beard:1997ic}. When we quote a Higgs mass value for $SU(3)$
from \eq{Higgs_CW} we set $\beta=4.50$.}
computed from \eq{Higgs_CW} is always much smaller then the
$Z^0$ mass, which is not phenomenologically acceptable. For example for
the parameters as in \fig{f_potential_SU3} we get $\cos(\theta_{\rm W})=0.82$
but $\rho_{HZ^0}=0.04$.
In order to get larger values of $\rho_{HZ^0}$
we have to lower $N_5$, as it is demonstrated by the data in
\tab{t_smallN5}.
\begin{table}[!t]
 \centering
  \begin{tabular}{lllllll}
   \hline\\
   \multicolumn{1}{c}{$N_5$}  &
   \multicolumn{1}{c}{$c$} &
   \multicolumn{1}{c}{$c_0$} &
   \multicolumn{1}{c}{$m_H R$} &
   \multicolumn{1}{c}{$\rho_{HZ^0}$} &
   \multicolumn{1}{c}{$\cos(\theta_{\rm W})$}&
   \multicolumn{1}{c}{$\alpha_{\rm min}$}\\[1.0ex]
   \hline\\
2 & 3.1  & 0.02  & 0.206 & 0.54 & 0.899 & 0.647\\[1.0ex]
2 & 50.0 & 0.2   & 0.828 & 1.25 & 0.888 & 0.655 \\[1.0ex]
3 & 3.1  & 0.0036& 0.095 & 0.23 & 0.868 & 0.641  \\[1.0ex] 
3 &50.0  & 0.05  & 0.400 & 0.43 & 0.836& 0.645  \\[1.0ex]
4 & 3.1  & 0.0007& 0.053 & 0.12 &0.877 & 0.645  \\[1.0ex]
4 & 50.0 & 0.01  & 0.238 & 0.21 &0.883 & 0.652  \\[1.0ex]
6 &3.1   & 0.0001& 0.021 & 0.04 &0.820& 0.638   \\[1.0ex] 
6 &50.0  & 0.0017& 0.123 & 0.07&0.888 & 0.653  \\[1.0ex]
   \hline
  \end{tabular}
 \caption{Results from the computation of the Coleman--Weinberg potential at
 small values of $N_5$. The Higgs mass is obtained from \eq{Higgs_CW} setting
 $\beta=4.50$. \label{t_smallN5}}
\end{table}

In fact for $N_5=2$, $c=50$ and $c_0=0.2$
we get $\rho_{HZ^0}=1.25$ which is the current lower limit from direct
searches for the Higgs boson. At this value of $N_5$ cut-off effects become
too large, the bound \eq{constraint} gives $\alpha<0.37$. Nevertheless
this is an indication that anisotropic lattices are needed to probe
the small $N_5$ region, where we might expect phenomenologically relevant
results.

\section{Conclusions}

Calculations of the Coleman--Weinberg potential for five-dimensional
gauge theories performed with an infinite cut-off yield the result of
absence of spontaneous symmetry breaking (SSB). Therefore the presence of
fermions is usually advocated.

In this work we have shown that SSB is observed in the pure gauge theory
when an explicit cut-off is introduced. The lattice was our choice, since
it is a gauge-invariant cut-off and we can compare the analytic results
to simulation data. The analytic results were produced using a Symanzik
effective langrangean, which is an expansion in operators of higher dimension
accompanied by powers of the lattice spacing. A truncation to the leading
order corrections is sufficient to produce evidence for SSB.
In \fig{f_mz_alpha} we show the non-perturbative spectrum of gauge bosons
for the $SU(2)$ theory compactified on an interval, as measured by
lattice simulations, and compare it with the Coleman--Weinberg calculation.
These results justify our approach. The presence of a Higgs phase with
a massive $U(1)$ gauge boson is expected from considerations based on
dimensional reduction. The Higgs boson in this theory has $U(1)$ charge
equal to 2 \cite{Irges:2006hg}. The four-dimensional Abelian Higgs model
for this charge has a phase separation between the confined and the Higgs phase
\cite{Fradkin:1978dv}.

We computed the Coleman--Weinberg potential for the phenomenologically more
relevant case of gauge group $SU(3)$ broken into $SU(2)\times U(1)$ by the
orbifold boundary conditions. We can reproduce the experimental value of the
Weinberg angle and a ratio of the Higgs mass over the $Z$ boson mass larger
than the current lower bound only if we lower the size $N_5$ of the extra
dimension in lattice units, which points at the use of anisotropic lattices.

Small $N_5$ with a fixed cut-off corresponds to a small $R$ and dimensional
reduction is expected to occur via compactification.  This would be the regime
where traditionally one would expect to see the SM. However, 
the large $N_5$ regime, which for a fixed cut-off corresponds to a large extra
dimension could also be interesting (large $N_5$ at fixed $R$ with large
cut-off drives us to the trivial point and we know what happens
there).\footnote{On the lattice this would correspond to 
choosing $N_5=L/a$.}
This would require some localization mechanism\footnote{
A prominent example is Ref. \cite{Dvali:1996xe}} to work and as argued in 
\cite{Irges:2006hg} such a situation seems to have chances to be realized near
the phase transition (small $\beta$). 
In addition, the towers of excited states collapse on their respective 
ground states \cite{Laine:2004ji}
pushing the KK expansion outside its domain of validity,
implying that the Coleman--Weinberg formula valid when the 
participating fields correspond to point particles, should not be trusted.
In fact, the Coleman--Weinberg potential due to the local excitations
is expected to be suppressed, see also the discussion in \cite{Green:2007tr}.
In this part of the phase diagram the global --- Polyakov loop --- nature
of the Higgs
and gauge bosons \cite{Irges:2006hg} can not be ignored. Beyond the lattice
very few systematic analytical approaches are known that can probe the theory
in this, inherently non-perturbative domain. The mean-field approximation 
could be one of them \cite{Knechtli:2005dw}.

%%%%%%%%%%%%%%%%%%%%%%%%%%%%

\bigskip

{\bf Acknowledgment.}
We would like to thank Philippe de Forcrand, Gero von Gersdorff and
Peter Weisz for helpful discussions. The computer time for the
lattice simulations has been kindly provided by 
the Swiss National Supercomputing Centre (CSCS) in Manno, by
the Institute for Theoretical Physics of the University of Berne and by
the University of Wuppertal.

\bigskip

%%%%%%%%%%%%%%%%%%%%%%%%%%%%
%\begin{appendix}
%\input{appa.tex}
%\end{appendix}
%%%%%%%%%%%%%%%%%%%%%%%%%%%%

\bibliography{SSB5D1}           %or whatever your .bib file is
\bibliographystyle{h-elsevier.bst}   %if you use h-elsevier.bst

%%%%%%%%%%%%%%%%%%%%%%%%%%%%
\end{document}